# Drive and measurement electrode patterns for electrode impedance tomography (EIT) imaging of neural activity in peripheral nerve

J. Hope[1,2], F. Vanholsbeeck[2], A. McDaid[1]

[1]Department of Mechanical Engineering, The University of Auckland, 5 Grafton Road, Auckland 1010, NZ
[2]Dodd Walls Centre, The Department of Physics, The University of Auckland, 38 Princes Street, Auckland 1010, NZ

**Abstract**
**Objective:** To establish the performance of several drive and measurement patterns in EIT imaging of neural activity in peripheral nerve, which involves large impedance changes in the nerve's anisotropic length axis.
**Approach:** Eight drive and measurement electrode patterns are compared using a finite element (FE) four-cylindrical shell model of a peripheral nerve and a 32 channel dual-ring nerve cuff. The central layer of the FE model contains impedance changes representative of neural activity of -0.30 in length axis and $-8.8 \times 10^{-4}$ in the radial axis. Four of the electrode patterns generate longitudinal drive current, which runs parallel to the anisotropic axis, while the remaining four patterns generate transverse drive current, which runs perpendicular to the anisotropic axis.
**Main results:** Transverse current patterns produce higher resolution than longitudinal current patterns but are also more susceptible to noise and errors, and exhibit poorer sensitivity to impedance changes in central sample locations. Three of the four longitudinal current patterns considered can reconstruct fascicle level impedance changes with up to 0.2 mV noise and error, which corresponds to between -5.5 and +0.18 dB of the normalised signal standard deviation. Reducing the spacing between the two electrode rings in all longitudinal current patterns reduced the signal to error ratio across all depth locations of the sample.
**Significance:** Electrode patterns which target the large impedance change in the anisotropic length axis can provide improved robustness against noise and errors, which is a critical step towards real time EIT imaging of neural activity in peripheral nerve.

## 1 Introduction

Electrical Impedance Tomography (EIT) is an imaging modality using electrical impedance as the contrast agent. Biomedical applications of EIT include monitoring of respiratory and pulmonary systems [2], identification of ischaemic brain tissue for diagnosis of stroke [3], and localisation of epileptic foci [4, 5] amongst others [6]. Differential EIT imaging of neural activity, 'neural-EIT', is made possible by transient changes in tissue impedance attributable to increased ion flux across neurone and axon membranes during neural activity [7]. The application of neural-EIT to a nerve cuff is an emerging field which offers a potential means to classify multiple concurrent compound action potentials within a peripheral nerve for neural prosthetics [1].

It is common in biological applications of EIT to configure the drive and measurement electrodes on the same plane in a ring around the sample boundary, which is then expanded to 3D EIT by stacking several rings, or 2D imaging planes [8-11]. Where stacked rings are not practical, such as half ellipsoid on the head, drive and measurement pattern strategies are employed to target a region of interest [12-16] and maximise information produced [17]. In a multi-contact nerve cuff the electrodes are configured in rings around the outer boundary of the nerve [18-20], which can be modelled as the surface of a cylinder [20]. EIT simulations by [21], for 3D EIT lung imaging of the thorax, compared seven possible drive EIT current and measurement patterns within a dual-ring electrode array comprised of two rings of eight electrodes. The authors modelled the lung and thorax as a cylindrical sample volume with isotropic conductivity features, and tested both longitudinal and transverse current patterns, concluding that transverse current patterns provided the best performance in presence of noise and electrode placement errors. The same seven electrode patterns were later tested in experiments on a cylindrical isotropic phantom and on the chests of 8 healthy adults in [22], concluding that an electrode pattern with a sequence of alternating transverse and longitudinal currents, called 'Square', provided the best performance.

The results in [21] and [22] assumed an isotropic sample, and so aren't directly transferrable to peripheral nerves because the region of interest for EIT, the nerve fibres, are highly aligned along the nerve length axis, producing electrical anisotropy. This fibre alignment, together with the fibre cytostructure, produces a fraction change in impedance during neural activity which is significantly higher in longitudinal direction, parallel to the fibres, than in the transverse direction, across fibres, as predicted by models of unmyelinated and myelinated fibres [1], and observed in in-vivo experiments [23]. Furthermore, the presence of anisotropy can produce boundary voltage data with non-unique solution [24], although numerical methods with some a-priori information have proven capable of reconstructing anisotropic anomalies in 2D simulations [24-26] and in 3D simulations to manage anisotropy of white matter in the brain [27, 28]. Transverse current patterns in a nerve would largely eliminate the anisotropy by operating in a plane perpendicular to the axis containing the unique conductivity (the 'anisotropic axis'), an approach adopted by [29] on peripheral nerve and by [30] on muscle tissue. However, because transverse current in a nerve is subject to a lower fraction change in impedance during neural activity, it suffers from a critically low signal to noise ratio [23]. Researchers typically circumnavigate this problem by averaging recordings across multiple measurements in order to reduce noise [17, 23, 29, 31], a practice which hinders real time neural-EIT. There is, therefore, a need to investigate whether the large impedance change in the longitudinal, anisotropic axis improves the signal to noise ratio of longitudinal current electrode patterns, and whether operating in the presence of anisotropy produces any detrimental effects on EIT reconstruction.

In this study, we present eight drive and measurement electrode patterns selected to contain either the maximum or minimum angular offset between electrode pairs. The electrode patterns are implemented, for the first time, on a finite element (FE) model of a four-shell cylindrical sample with anisotropic conductivity along the length axis of the central shell. Performance is evaluated using several quantitative metrics as well as a qualitative analysis of the reconstructed conductivity map. The results highlight the influence that the anisotropic impedance characteristics of the neural environment have on the performance of electrode patterns. The study is designed with specific interest in neural-EIT of peripheral nerves, although the results are relevant to any cylindrical sample with an anisotropic length axis e.g. muscle.

## 2 Methods

*A. EIT Forward Solution*

Solutions to the EIT forward problem were obtained using a cylindrical shell model approximation of a 50mm length, single fascicle, sciatic nerve of rat as described in [1], Table 1. The intra-fascicle tissue was divided into a grid of 49 sub-volumes each extending the length of the nerve, Fig 1a. The nerve cuff with dual-ring electrode array contained two rings of 16 electrodes, each 1.1 x 0.11mm in size, arranged into on a 22.5° pitch around the circumference. Electrodes were implemented with the complete electrode model (CEM) [32], including a contact impedance of $1.5 \times 10^{-4}$ $\Omega m^2$ at the electrode-saline interface and an electrode conductivity of $4 \times 10^6$ S/m. Simulations were run with the two electrodes rings spaced at 3mm, 6mm and 10mm apart along the nerve length. All external surfaces were insulated.

**Table 1:** FE model parameters for single fascicle model [1]. 'T' = Transverse, 'L' = Longitudinal.

| Layer | Radius (µm) | Conductivity (S/m) |
| --- | --- | --- |
| Intra-fascicle tissue | 545 | Inactive: T: 0.325 / L: 0.08757<br>Active: T: 0.4348 / L: 0.08764 |
| Perineurium | 550 | 0.021 |
| Epineurium | 600 | 0.08257 |
| Saline | 680 | 2 |

The FE model was implemented in COMSOL Multiphysics version 5.3, using Electric Currents physics in the AC/DC module, on a Dell Optiplex 7040 PC with Intel i7-6700 processor. Meshing was performed with minimum mesh size of 0.1mm, a max growth rate of 1.3, a curvature factor of 0.2, and a resolution of narrow regions of 1, producing a total of 750k free tetrahedral elements with minimum and average mesh quality of $2.8 \times 10^{-2}$ and 0.54 respectively. The FE model was then solved for each drive and measurement pattern with a 10µA amplitude [23] drive current applied between each of the 16 drive electrode pairs one pair at a time, with the remaining electrodes acquiring boundary voltage measurements. A quasi-static approximation of Maxwell's equations was used to implement neural activity by solving the FEM models under several static conditions, where, for each condition, the electrical conductivity of fascicle sub-volumes was set to that of either the active or inactive state. This quasi-static approach is valid up to several 100s of kHz for intra-fascicle tissue [1].

*B. EIT Inverse Solution*

Zeroth order Tikhonov regularisation was used to invert the sensitivity matrix, as was done in [17, 33], with the Tikhonov regularisation parameter selected using the L-curve, or Pareto frontier curve, method [34]. To replicate real operating conditions, we synthesised data in line with recommendations provided in [32]; these are: 1) to use a smaller mesh size: minimum 0.05 mm, resulting in 4.3M free tetrahedral elements with minimum and average mesh quality of $7.0 \times 10^{-2}$ and 0.60 respectively; 2) ensure different size, shape and location of conductivity features: 3 fascicle model with 90 µm thick epineurium, Fig 1b; 3) add either low (+/-0.35 $\mu V_{RMS}$) or high (+/- 35$\mu V_{RMS}$) Gaussian noise to each measurement:; 4) add random hardware error to each electrode (channel): within either a low (+/- 1 µV) or high (+/- 100 µV) range; and 5) quantization: either low (1 µV) or high (10 µV) rounding. All low noise, error and quantization conditions were applied together, as were all high conditions.

Noise and errors were selected as absolute values which are independent of the magnitude of the boundary voltages. In practice, some error sources are dependent on the signal size, such as noise in the drive current and electrode impedance measurement errors, whereas others are independent of signal size, such as noise and accuracy in the ADC hardware.

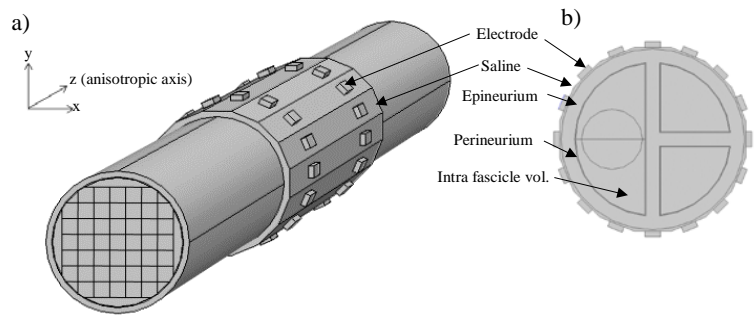

**Figure 1:** 3D end-and-side view of the single fascicle model used for the EIT forward problem (a), and end view of the 3 fascicle model used to generate data for the EIT inverse problem (b).

*C. Drive and Measurements Patterns*

Eight drive and measurement patterns were investigated: four utilising longitudinal current, implemented in a dual ring electrode array, and four utilising transverse current, implemented in a single-ring electrode array. In each pattern, current flows between the drive electrode pair, and the remaining electrodes are paired up to produce differential measurements. The drive electrode pairs were selected with either the maximum or minimum angular offset between electrode pairs, which, respectively, minimises or maximises current channelling through the low resistance outer fluid layer [35]. The measurement electrode pairs were then configured to either match or mismatch the drive pair.

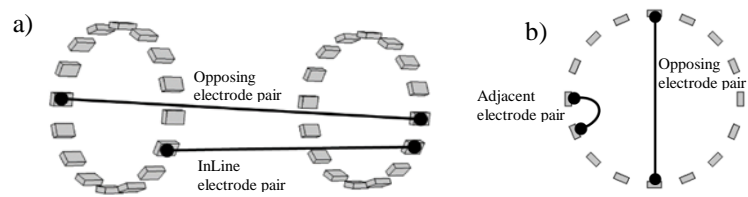

**Figure 2:** Angled view of the dual-ring electrode array with, labelled, a pair of opposing and a pair of in-line drive or measurement electrodes used with longitudinal drive currents (a), and an end-on view of a single-electrode ring with, labelled, a pair of opposing and a pair of adjacent drive or measurement electrodes used with transverse drive currents (b).

With longitudinal current, drive and measurement electrode pairs were on different electrode rings and were either 'Opposing', i.e. at an angular offset of 180 degrees, or 'InLine', i.e. an angular offset of 0 degrees, Fig 2a. We combined the longitudinal drive and measurement patterns following the convention in [21] to produce two possible patterns (drive/measurement): InLine/InLine and Opposing/Opposing as well as in a non-conventional manner to produce two further patterns: Opposing/InLine and InLine/Opposing. Of these longitudinal current configurations Opposing/Opposing featured in studies by Refs [21, 22] under the name 'Zigzag-opposite'.

With Transverse current, drive and measurement electrode pairs were on the same ring and were either 'Opposing', i.e. 180 degrees angular offset, or 'Adjacent', i.e. between two neighbouring electrodes, Fig 2b. Transverse: Opposing/Opposing and Transverse: Adjacent/Adjacent patterns were both evaluated in Refs [21, 22] under the names 'Planar-opposite' and 'Planar' respectively.

Opposing, In-Line, and Adjacent are hereon abbreviated to O, I, and A respectively when referring to drive/measurement pairings.

*D. Performance metrics*

Electrode patterns were compared using three criteria: 1) analysis of the condition number together with singular values from the singular value decomposition (SVD) of the sensitivity matrix; 2) the signal to error ratio, which we define as:

$$SER = 20 \log(v_\sigma/v_e) \quad (1)$$

where $v_\sigma$ is the standard deviation of the normalised differential boundary voltage measurements obtained from all 16 drive electrode pair conditions, and $v_e$ is the maximum possible voltage error from noise and hardware errors normalised using the mean of differential measurements from the inactive state; 3) qualitative analysis of the reconstructed conductivity map, where poor quality reconstruction includes conductivity changes in wrong locations and/or significantly smaller in magnitude.

**3 Results**

Inspection of the sensitivity matrix for Transverse: O/O pattern revealed normalised boundary voltage measurements were several orders of magnitudes larger on the electrode pair at +/- 90 degrees positions relative to the drive current electrode pair (i.e. on the left and right hand sides in Fig 2b) due to symmetry of the circular sample cross section. Differential boundary voltages of negligible magnitude produced a negligible mean value, which in turn distorts the normalised boundary voltages through its calculation as the difference divided by the mean. Measurements from this electrode pair were therefore excluded from the sensitivity matrix.

**Table 2:** Condition number of the sensitivity matrix for each drive and measurement electrode configuration.

| Current: drive/measurement pattern | N/A | Electrode ring spacing | | |
|---|---|---|---|---|
| | | 3 mm | 6 mm | 10 mm |
| Longitudinal: O/O | | 3.79E+6 | 7.18E+6 | 1.36E+7 |
| Longitudinal: O/I | | 2.03E+6 | 4.33E+6 | 1.84E+6 |
| Longitudinal: I/I | | 1.57E+6 | 3.07E+6 | 1.54E+6 |
| Longitudinal: I/O | | 4.45E+6 | 6.02E+6 | 4.01E+6 |
| Transverse: O/O | 4.57E+5 | | | |
| Transverse: O/A | 1.71e+5 | | | |
| Transverse: A/A | 3.04E+4 | | | |
| Transverse: A/O | 3.25E+5 | | | |

Longitudinal patterns exhibited minimal difference in condition numbers, where large condition numbers indicate a more ill-conditioned sensitivity matrix, and in singular values, Table 2 and Fig 3 respectively. The same was true of transverse patterns with the exception of Transverse: O/O, where the singular values were considerably lower than the other transverse patterns. In comparing longitudinal and transverse patterns: condition numbers were consistently higher and singular values were consistently lower in longitudinal patterns than in transverse patterns, albeit to a lesser extent with Transverse: O/O. Transverse patterns appear to be, as a whole, more robust against errors relative to the signal size.

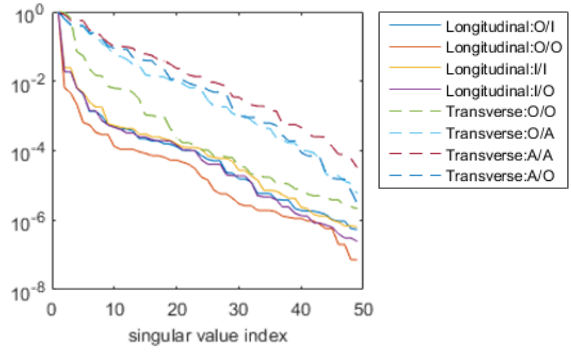

**Figure 3:** Normalised singular values, from singular value decomposition of the sensitivity matrix, for each electrode pattern, with 6mm ring spacing in longitudinal current patterns.

A large positive SER corresponds to high robustness against noise and error, whereas negative SERs indicate the signal standard deviation is below the specified noise and error level. With longitudinal patterns the signal standard deviation $v_\sigma$, from equation 1, increased with decreasing ring spacing, as did $v_e$. The net effect was a decrease in SERs across all grid indices as ring spacing decreased, Fig 4a. With transverse patterns, O/O produced the highest SER across all grid indices, and A/A the worst, Fig 4b. In comparing longitudinal and transverse patterns: the troughs in SER plots, with minima corresponding to central grid indices 9, 16, 23, 30 and 37, Fig 4c, are significantly larger in all transverse patterns than in all longitudinal patterns, indicating that longitudinal patterns provide better sensitivity to conductivity changes in the centre of the sample. Similarly, with the quarter circle shaped fascicle active in the multi-fascicle model, a high noise and error level of 0.2 mV produced SERs of between -5.5 and +0.18 dB in longitudinal patterns with 10mm electrode spacing, and between -46 and -19 dB in transverse patterns. The higher SERs of longitudinal patterns indicate they are more robust than transverse patterns against noise and error in voltage measurements.

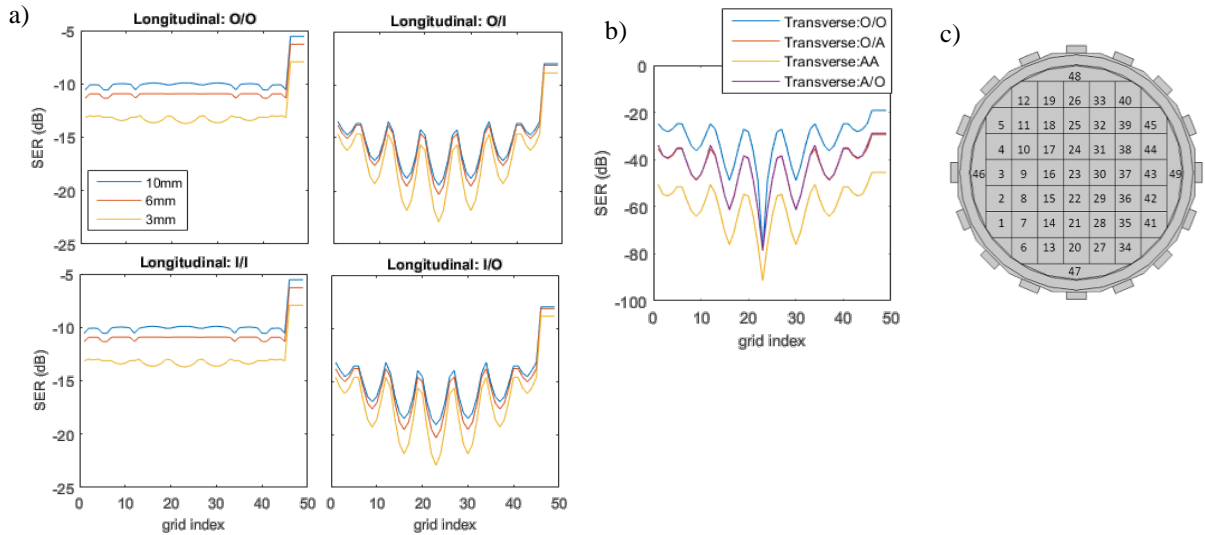

**Figure 4:** The signal to error ratio (SER) at each grid index for longitudinal electrode patterns of 10mm, 6mm and 3mm ring spacing (a), and for transverse current patterns (b). The grid in the single fascicle nerve model are shown overlaid with the corresponding grid index in (c).

Under the low noise and error conditions, EIT reconstruction of impedance changes in two quarter shaped fascicles and in one quarter-circle shaped fascicle could be distinguished from one another in all electrode patterns, Figs 5a and 5b. Under the high noise and error conditions, longitudinal patterns remained relatively unchanged, whereas the transverse patterns were severely affected, as exhibited by significant impedance changes in significantly wrong locations, Fig 5b. Where noise did not corrupt the results, O/O pattern in both longitudinal and transverse currents reconstructed a "mirrored" impedance distribution, such that equal magnitude impedance changes in grids on the both the correct side and opposing side of the nerve cross section were observed. The other three transverse patterns and Longitudinal: I/I also produced impedance changes in grids in significantly wrong locations, although often with far smaller impedance magnitude than those grids corresponding to the correct locations. Overall, with the exception of the two O/O patterns, the transverse patterns appear to offer higher resolution than the longitudinal patterns, as exhibited by tighter clusters of grids with impedance change, but also a higher susceptibility to noise and errors which manifests as impedance changes in significantly wrong locations.

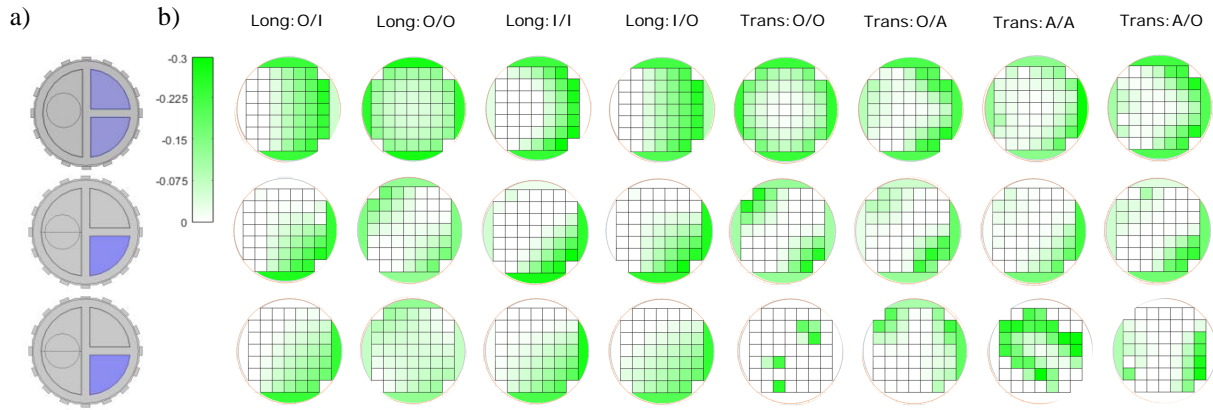

**Figure 5:** Activity in the three fascicle model with purple indicating -0.3 and -8.8x10$^{-4}$ fraction change in longitudinal and transverse impedances respectively (a), and the corresponding EIT reconstruction for each electrode pattern with low noise and error conditions (top and middle rows) and high noise and error conditions (bottom row) (b). Colour scale shows the reconstructed fraction change in impedance. Longitudinal drive patterns used 10mm electrode spacing.

## 4 Discussion

Analysis of condition numbers and singular values predicted transverse patterns to be, as a whole, more robust against errors relative to the signal size, i.e. the normalised boundary voltages. SER analysis and visual analysis of reconstructions both indicated that longitudinal patterns are more robust than transverse patterns against an absolute value of noise and errors, particularly in central sample locations, but offer poorer resolution. This is contrary to modelling in [21] on an isotropic sample, which predicted transverse patterns to be more robust against noise and errors. Therefore, in nerves, targeting the large fraction change in impedance in the longitudinal axis using longitudinal current outweighs the poorer performance previously observed in longitudinal patterns. The recommendation by [22] to use a 'Square' pattern, which combines aspects of Transverse: A/A and Longitudinal: I/I, is an interesting approach, and appears to offer the benefits of both transverse and longitudinal current patterns. However, the demanding time resolution requirements of imaging neural activity may restrict implementation of additional drive electrode pairs as all pairs must be cycled through at least once within each time resolution increment.

Neural-EIT researchers who have utilised transverse patterns have typically averaged recordings across multiple measurements in order to reduce noise [17, 23, 29, 31]. The significantly improved SER of longitudinal patterns may remove this need for averaging data and, therefore, open up the possibility of imaging neural activity in real time. Other factors to be considered when selecting between transverse and longitudinal patterns for neural EIT include: 1) the excitation threshold of fibres is higher in transverse orientation [36], potentially allowing a higher drive current amplitude; 2) the transmembrane capacitive charge transfer mechanism occurs at higher frequencies in transverse fibre orientation, allowing higher drive current frequencies [1].

All longitudinal patterns exhibited a trend of decreasing SER with decreasing electrode spacing, albeit only marginally with I/O and O/I patterns. This trend is expected to be countered by an increase in the fraction change in impedance during neural activity as electrode spacing decreases, which is predicted by modelling [1] but has not been included here in order to isolate the performance of electrode patterns from this system variable. The larger increase in $v_e$ than $v_\sigma$ produces the negative trend between SER and ring spacing, and is, in part, an implication from selecting a noise value which is independent of the ring spacing. Noise sources which are dependent on the ring spacing, such as noise in the current source, would counter the observed trend.

An opposing drive current is desirable as it is reported to minimise current channelling through low resistance outer layers [35]. However, for both longitudinal and transverse currents O/O pattern produced significant errors in the EIT reconstruction, which is in agreement with observations in [21] and [22] on opposing electrode patterns - although a direct comparison is not possible in the longitudinal current case due to anisotropy in our sample. Opposing electrode patterns did not display the same susceptibility to errors when coupled, in a non-conventional manner, with a different (InLine or Adjacent) drive electrode pattern in either the drive or measurements electrode pairs.

*A. Limitations*

The model includes realistic drive current amplitude, noise levels, hardware errors, quantization, different mesh sizes and different conductivity distributions. However, inaccuracies are generated from the assumption of a homogeneous intra-fascicle tissue medium, which neglects to consider that the true direction of the current path is restricted by the cytostructure of nerve fibres, for example to enter and exit through nodes of Ranvier. The geometric assumptions, such as perfect alignment of the two electrode rings and perfectly centred nerve within the cuff, would be difficult to replicate in practice. The fascicle and nerve dimensions are chosen to approximate those of the rat sciatic nerve which allows comparison to in-vivo animal studies [29, 37] but must be scaled up in number and size, respectively, to approximate major peripheral nerves, e.g. the medial nerve, in human. The impedance change applied to intra-fascicle tissue, Table 1, is representative of all fibres being active [1], and so provides an artificially large signal. To accommodate the large signal, the absolute noise and error value of 0.2 mV is selected to be a factor of 20 to 200 times greater than that commonly found in neural EIT systems [7, 17, 31]. Insulation of the outer surfaces in the FE models, as opposed to an infinite boundary, artificially increases the SERs. This effect on the SERs is minimal, < 0.2 dB at each grid index, but without it, a reduction in resolution of the reconstructions is expected.

As a whole, the accuracy of the model presented here is sufficient to compare performance of electrode configurations in reconstructing impedance changes in a highly anisotropic length axis, such as in peripheral nerve.

**5 Conclusion**

We have identified three longitudinal current patterns, InLine/InLine (I/I), Opposing/InLine (O/I) and InLine/Opposing (I/O), which are more robust against noise and errors than their transverse current counterparts, and exhibit no detrimental effects from operating in the anisotropic length axis. The large SER is a critical step towards real time imaging of neural activity using EIT, where the current practice of averaging multiple measurements to improve the SER is not practical.